\documentstyle[prd,aps,floats,epsf]{revtex}

\def\be{\begin{equation}}
\def\ee{\end{equation}}

\def\ea{\end{eqnarray}}

\begin{document} 
\preprint{BROWN-HET-1142, hep-ph/yymmddd}
\draft

\renewcommand{\topfraction}{0.99}
\renewcommand{\bottomfraction}{0.99}
\twocolumn[\hsize\textwidth\columnwidth\hsize\csname 
@twocolumnfalse\endcsname

\title
{\Large {\bf Adiabatic and Isocurvature Perturbations for Multifield
Generalized Einstein Models}}

\author{F. Di Marco $^1$\footnote{e-mail: dimarco@bo.infn.it}, F. Finelli 
$^2$\footnote{e-mail: finelli@tesre.bo.cnr.it} and
R. Brandenberger $^3$\footnote{e-mail: rhb@het.brown.edu}
\\
$^1$ Dipartimento di Fisica, Universit\`a degli Studi di Bologna
and I.N.F.N., \\ via Irnerio, 46 -- 40126 Bologna -- Italy \\
$^2$ I.A.S.F. - Sezione di Bologna, C.N.R., Via Gobetti 101,
40129 Bologna, Italy,\\
$^3$ Department of Physics, Brown University, Providence, RI 02912, USA}

\date{\today}
\maketitle

\begin{abstract}
Low energy effective field theories motivated by string theory will
likely contain several scalar moduli fields which will be relevant to
early Universe cosmology. Some of these fields are expected to
couple with non-standard kinetic terms to gravity. In this paper,
we study the splitting into adiabatic and isocurvature perturbations for a
model with two scalar fields, one of which has a non-standard kinetic
term in the Einstein-frame action.  Such actions 
can arise, e.g., in the Pre-Big-Bang and Ekpyrotic scenarios. The
presence of a non-standard kinetic term induces a new coupling between
adiabatic and isocurvature perturbations which is non-vanishing when 
the potential for the matter fields is nonzero.
This coupling is un-suppressed in the long wavelength limit and thus
can lead to an important transfer of power from the entropy to the adiabatic
mode on super-Hubble scales.
We apply the formalism to the case of a
previously found exact solution with an exponential potential and study
the resulting mixing of adiabatic and isocurvature fluctuations in this
example. 
We also discuss the possible relevance of the extra coupling in the 
perturbation equations for the process of generating 
an adiabatic component of the fluctuations spectrum from isocurvature 
perturbations without considering a later decay of the isocurvature
component.
\end{abstract}

\pacs{PACS numbers: 98.80Cq}]

\vskip 0.4cm

\section{Introduction}

There has been a lot of recent interest in cosmological models motivated
by string theory, in particular in models in which the dynamics differs
from that of standard scalar field-driven inflationary models. Examples
of such models include the ``Pre-Big-Bang (PBB)'' scenario
\cite{Veneziano:1991ek,Gasperini:1992em,Lidsey:1999mc,Gasperini:2002}, the
Ekpyrotic model \cite{Khoury:2001wf}, and ``Mirage Cosmology'' 
\cite{Kraus:1999it,Kehagias:1999vr,Alexander:1999cb}.

The simplest realizations of the PBB and the Ekpyrotic scenarios as
four space-time dimensional effective field theories involve (in the
Einstein frame) Einstein gravity coupled minimally to a single scalar
field with canonical kinetic term. This scalar field can be
viewed as the conformally rescaled dilaton in the case of PBB cosmology
or the position modulus of a bulk brane in Ekpyrotic cosmology. These
realizations, however, fail to produce a scale-invariant ($n = 1$)
spectrum of cosmological fluctuations, producing instead an $n = 4$ blue
spectrum in the case of PBB cosmology \cite{Brustein:1994kn}, and an $n = 3$
spectrum in the case of Ekpyrotic cosmology 
\cite{Lyth:2001pf,Brandenberger:2001bs,Hwang:2001ga,Tsujikawa:2001ad,Tsujikawa:2002qc}
(although the latter conclusion is still not agreed on, the works of
\cite{Khoury:2001zk,Durrer:2002jn} claiming to obtain an $n = 1$ spectrum).

String-motivated early Universe scenarios, in particular the PBB and
Ekpyrotic cosmologies, start from basic physics which, in addition to
the degrees of freedom mentioned above, contain scalar fields
with non-standard kinetic tersms, for example the axion 
\cite{Copeland:1997ug} in the case of both PBB 
and Ekpyrotic cosmologies, which may play an important role in the
early Universe. In fact, in the context of Ekpyrotic cosmology it may
not be consistent to neglect these fields \cite{Rasanen:2001hf,Kallosh:2001ai}.

More generally, string theory contains many scalar moduli fields
which can be expected to be important in early Universe cosmology.
Independent of which frame one uses for computations, some of the
moduli fields will have non-standard kinetic terms in the action.
Hence, it is important for future applications to string cosmology
to provide a framework for studying the joint metric and matter
fluctuations in such models.

The presence of more than one scalar matter field leads to the
existence of entropy modes in the spectrum of cosmological fluctuations
which are important on large scales (length scales greater than the
Hubble radius). It has been suggested that such entropy modes might
lead to an acceptable (i.e. nearly scale-invariant) spectrum of
fluctuations in PBB and Ekpyrotic cosmologies. In the case of PBB
cosmology, this was worked out several years ago \cite{Copeland:1997ug},
resulting in the possibility of a scale-invariant spectrum of
fluctuations which is of primordial isocurvature nature, and it has
recently been pointed out \cite{Bozza:2002fp} that it is possible to
convert this spectrum into a scale-invariant adiabatic spectrum
if the axion is short-lived (making use of the ``curvaton'' 
\cite{Lyth:2001nq,Mollerach:hu,Enqvist:2001zp,Moroi:2001ct} 
mechanism). In the case of Ekpyrotic
cosmology, the possibility of obtaining a scale-invariant
spectrum of fluctuations making use of a second scalar field was
recently explored in \cite{Notari:2002yc,assicontr} (see
\cite{Brandenberger:2001bs} for an earlier paper drawing attention to this
possibility). 

To be able to predict the scaling of cosmological fluctuations in
models with more than one scalar matter field, it is important to
understand the coupling between the adiabatic and the entropy
modes. It has been known for a long time (see e.g. \cite{Axenides:1983hj}
for a study of axion fluctuations in inflationary cosmology) that
an initial entropy fluctuation immediately begins to source the
growing adiabatic mode, even on scales larger than the Hubble radius. 
In the case of two minimally coupled scalar fields, a 
convenient new formalism to study both adiabatic and
entropy fluctuations was recently developed by Gordon et al.
\cite{Gordon:2000hv}. In this paper, we generalize this formalism to the
case of relevance in many models of string cosmology in which one of the 
scalar fields (e.g. the axion) has a non-canonical kinetic term,
in the sense that its kinetic term in the total Lagrangian depends
on the value of the first scalar field (e.g. the dilaton).  

We consider the following action:
\be
S = \int d^4 x \sqrt{-g} \left[ \frac{R}{16 \pi G} - \frac{1}{2}
(\partial \varphi)^2 - \frac{e^{2 b (\varphi)}}{2}
(\partial \chi)^2 - V (\varphi, \chi) \right]
\label{action}
\ee
where 
$8 \pi G = M_{\rm pl}^{-2}$. Such an action is motivated by various
generalized Einstein theories \cite{Berkin:nm,Starobinsky:2001xq} and 
occurs in the case of PBB cosmology with $\chi$ as an axion field. It is
also of interest in Ekpyrotic cosmology, since it is a two field
simplification of the 4D effective action of
low-energy heterotic M-theory \cite{Copeland}. However, the action
(\ref{action})
contains a potential, which is crucial in our analysys of these string
motivated cosmologies. Infact, in appendix A, we show that the most
general free 4D effective action derived from M-theory with a moving
brane, cannot lead to a scale invariant perturbation neither in the
adiabatic nor in the isocurvature sector. 


We find that (in the case of non-vanishing potential) 
the presence of a non-canonical kinetic term
for $\chi$ leads to an extra term in the equations of motion
which couple adiabatic and isocurvature fluctuation modes on
wavelengths larger than the Hubble radius. This
coupling term does not vanish even when the ratio of the kinetic
energies of the two matter fields is constant ($\theta$ constant
in the notation introduced in Section 2). The nonvanishing of the
coupling can lead to a more rapid growth of the curvature fluctuation
generated by an isocurvature perturbation.

The outline of the paper is as follows. In Section II we present the
formalism of the average field $\sigma$ and the orthogonal field $s$ for
the background of action (\ref{action}). Section III collects all the
relevant formulae for adiabatic and isocurvature perturbations. These two
sections generalize the work of Gordon et al. \cite{Gordon:2000hv}
for minimally coupled scalar fields with standard kinetic terms .
In Sections IV and V we discuss cosmological perturbations around an exact
solution previously found in \cite{Finelli:2001sr}, and  
Section VI summarizes our main conclusions. 
In Appendix A, we discuss the cosmological perturbations for a 
free 4D effective action deriving from heterotic M-theory with a moving
five brane. In Appendix B, we discuss the stability 
of the background solution found
in \cite{Finelli:2001sr}. 

\section{Background}

The equations of motion for the fields $\varphi$ and $\chi$ which follow
from the action (\ref{action}) are:
\be
\ddot \varphi + 3 H \dot \varphi + V_\varphi = b_\varphi
e^{2 b} \dot \chi^2 \,,
\label{backphi}
\ee
\be
\ddot \chi + (3 H + 2 b_\varphi \dot \varphi) \dot \chi 
+ e^{- 2 b} \, V_\chi = 0 \,,
\label{backchi}
\ee
where we denote the derivative with respect to a field by the
corresponding subscript. The Einstein equations are: 
\be
H^2 = \frac{1}{3 M_{\rm pl}^2} \left[ \frac{1}{2} \dot \varphi^2 + 
\frac{e^{2 b}}{2} \dot \chi^2 + V \right] \,,
\ee
and
\be
\dot H = - \frac{1}{2 M_{\rm pl}^2} \left[ \dot \varphi^2 +
e^{2 b} \dot \chi^2
\right] \,. 
\ee

We now use the formalism of average and orthogonal field, extending the
formalism developed by Gordon et al. \cite{Gordon:2000hv} to the case with
non-canonical kinetic terms. 
The definition of the time derivative of the average field
$\sigma$ is:
\be
d \sigma = \cos \theta \, d \varphi + \sin \theta
\, e^b \, d \chi
\label{adiabaticfield}
\ee
where 
\begin{eqnarray}
\cos \theta &=& \frac{\dot \varphi}{\sqrt{\dot \varphi^2 +
e^{2 b} \dot \chi^2}} \nonumber \\
\sin \theta &=& \frac{e^b \dot \chi}{\sqrt{\dot \varphi^2 +
e^{2 b} \dot \chi^2}} \,.
\end{eqnarray}
These equations generalize Eqs. (37-38) of Ref. \cite{Finelli:2001sr}. 
It is evident that
\be
\dot \sigma = \sqrt{\dot \varphi^2 + e^{2 b} \dot \chi^2}
\ee 
and that the average field $\sigma$ satisfies
\be
\ddot \sigma + 3 H \dot \sigma + V_{\sigma} = 0
\ee
where 
\be
V_{\sigma} = \cos \theta \, V_\varphi + e^{- b} \sin \theta
\, V_\chi \, .
\ee

\section{Perturbations}

We now write the equation for the field perturbations. We will expand to
first order the equations of motion for $\varphi$ and for $\chi$:
\begin{eqnarray}
- g^{\mu \nu} \nabla_\mu ( \partial_\nu \varphi ) + V_\varphi +
b_\varphi e^{2 b} g^{\mu \nu}
\partial_\mu \chi
\partial_\nu \chi &=& 0
\nonumber \\
- e^{-2 b} g^{\mu \nu} \nabla_\mu ( e^{2 b} \partial_\nu \chi ) +
e^{-2 b} V_\chi &=& 0 
\end{eqnarray}

In the longitudinal gauge (see e.g. Ref. \cite{Mukhanov:1990me}
for a detailed review of the theory of cosmological perturbations) 
and in the case of vanishing anisotropic stress (as is the case for
the matter Lagrangians considered in this paper) the metric perturbations
can be written as follows:
\be
ds^2 = - ( 1 + 2 \Phi) dt^2 + a^2 (1 - 2 \Phi) d {\bf x}^2 \, ,
\ee
where $\Phi$ is a function of space and time representing the metric
fluctuations. 

In this gauge, the fluctuation equation for the matter field
$\varphi$ becomes:
\begin{eqnarray}
\ddot \delta \varphi &+& 3 H \dot \delta \varphi + 
 \left[ \frac{k^2}{a^2} + V_{\varphi \varphi} 
- (b_{\varphi \varphi} + 2 b_\varphi^2) \dot \chi^2 e^{2 b} \right] 
\delta \varphi \nonumber \\ && \,\, +
V_{\phi \chi} \delta \chi - 2 b_\varphi e^{2 b} \dot \chi \dot \delta \chi 
\nonumber \\
&=& \dot \varphi ( 4 \dot \Phi + 3 H \Phi ) + (2 \ddot \varphi + 3 H \dot 
\varphi) \Phi - 2 b_\varphi e^{2 b} \dot \chi^2 \Phi
\nonumber \\
&=& 4 \dot \varphi \dot \Phi - 2 V_\varphi \Phi \, ,
\label{deltaphi}
\end{eqnarray}
where in the second line we have used the equation of motion for the
background (\ref{backphi}). 
Similarly, the equation of motion for the 
fluctuation in the $\chi$ field can be written as:
\begin{eqnarray}
\ddot \delta \chi &+& (3 H + 2 b_\varphi \dot \varphi) \dot
\delta \chi + 
\left[ \frac{k^2}{a^2} + e^{- 2 b} V_{\chi \chi} \right] \delta \chi 
+ 2 b_\varphi \dot \chi \, \dot \delta \varphi
\nonumber \\
&& \,\, + e^{- 2 b} \left[ V_{\chi \varphi} - 2 b_\varphi V_\chi + 
2 b_{\varphi \varphi} \dot \varphi \dot \chi \right] \delta \varphi 
\nonumber \\
&=& \dot \chi (4 \dot \Phi + 3 H \Phi ) + (2 \ddot \chi + 3 H \dot
\chi) \Phi + 4 b_\varphi \dot \varphi \dot \chi
\Phi \nonumber \\
&=& 4 \dot \chi \dot \Phi - 2 e^{- 2 b} V_\chi \Phi \, .
\label{deltachi}
\end{eqnarray}
Note that these two last two equations agree with those in 
\cite{Garcia-Bellido:1995fz}.

The energy and the momentum constraint equations are, respectively:
\begin{eqnarray}
&& 3 H (\dot \Phi + H \Phi) + \dot H \Phi  + \frac{k^2}{a^2} \Phi = \\
&-& \frac{1}{2 M_{\rm pl}^2} 
\left[ \dot \varphi \dot \delta \varphi + e^{2 b} \dot \chi
\dot \delta \chi + b_\varphi e^{2 b} \dot \chi^2 \delta \varphi
+ V_\varphi \delta \varphi + V_\chi \delta \chi \right] \nonumber
\end{eqnarray}
\be
\dot \Phi + H \Phi = \frac{1}{2 M_{\rm pl}^2} \left( \dot \varphi 
\delta \varphi + e^{2 b} \dot \chi \delta \chi \right) = 
\frac{1}{2 M_{\rm pl}^2} \dot \sigma\delta \sigma \, ,
\ee
and the redundant equation of motion for $\Phi$ is:
\begin{eqnarray}
&& \ddot \Phi + 4 H \dot \Phi + (\dot H + 3 H^2) \Phi = \label{Phisecond} \\
&& \frac{1}{2 M_{\rm pl}^2} \left[ \dot \varphi \dot \delta \varphi 
+ e^{2 b} \dot \chi \dot \delta \chi +
b_\varphi e^{2 b} \dot
\chi^2 \delta \varphi - V_\varphi \delta \varphi - V_\chi \delta \chi
\right] \, . \nonumber
\end{eqnarray}

Crucial for the following is the separation of the fluctuations into
the adiabatic and isocurvature modes.
The adiabatic fluctuation is associated with the ``average'' field
defined in Eq. (\ref{adiabaticfield}). 
The {\em entropy} field $s$ is given by
\cite{Gordon:2000hv,Notari:2002yc}:
\be
ds = \frac{e^b}{\dot \sigma} 
[ \, \dot \varphi \, d \chi - \dot \chi \, d \varphi \, ] = 
e^b \cos \theta \, d \chi 
- \sin \theta \, d \varphi 
\label{entropyfield}
\ee
(with the $ds$ not to be confused with the metric).

The curvature perturbation in comoving gauge
is \cite{Mukhanov:1990me,Lyth:1984gv}
\begin{eqnarray}
\zeta &=& \Phi -\frac{H}{\dot H} \left( \dot \Phi + H \Phi\right) 
\nonumber \\
&=& \Phi + H \left( \frac{\dot \varphi \delta \varphi + 
e^{2 b} \dot \chi \delta \chi}
{\dot \varphi^2 + e^{2 b} \dot \chi^2} \right) \, ,
\end{eqnarray}
and its evolution equation is (as follows from Eq. (\ref{Phisecond}))
\begin{eqnarray}
\dot \zeta &=& \frac{k^2}{a^2} \frac{H}{\dot H} \Phi + S \nonumber \\
&\equiv& \frac{k^2}{a^2} \frac{H}{\dot H} \Phi - H 
\left[\frac{1}{2} \frac{d}{dt} 
\left( \frac{e^{2b} \dot \chi^2 - \dot \varphi^2
}{ e^{2b} \dot \chi^2 + \dot \varphi^2}\right) 
+ \dot C \right] \left( \frac{\delta \varphi}{\dot \varphi} - 
\frac{\delta \chi}{\dot \chi} \right) \nonumber \\
&=& \frac{k^2}{a^2} \frac{H}{\dot H} \Phi - 2 H 
\frac{V_\varphi \dot \varphi \dot \chi^2 e^{2b}
- V_\chi    \dot \chi  \dot \varphi^2  }
{(e^{2b} \dot \chi^2 + \dot \varphi^2)^2} 
\left( \frac{\delta \varphi}{\dot \varphi} - 
\frac{\delta \chi}{\dot \chi} \right) \, , \nonumber \\
\label{correct}
\end{eqnarray}
where
\be
\dot C = 2 b_\varphi \frac{ \dot \varphi \, \dot \chi^2 e^{2b}}
{e^{2b} \dot \chi^2 + \dot \varphi^2} = 2 b_\varphi 
\dot \varphi \sin^2 \theta \,.
\ee
We note that Eq. (\ref{correct}) 
corrects Eq. (4.8) of \cite{Garcia-Bellido:1995fz}. Indeed,
in the free field case (vanishing potential), the coefficient multiplying
$\delta \varphi/ \dot \varphi - \delta \chi / \dot \chi$ should vanish,
and this does not occur in Eq. (4.8) of \cite{Garcia-Bellido:1995fz}.

We can also express the variation of $\zeta$ in Eq. (\ref{correct})
in terms of $\delta s$:
\begin{equation}
S = 2 \frac{H}{\dot \sigma} \dot
\theta \delta s + 2 b_\varphi H \sin \theta \delta s
= - 2 \frac{H}{\dot \sigma^2} V_s \delta s \,, 
\label{source}
\end{equation}
and we have used the relations \cite{Notari:2002yc}:
\be
\dot \theta = \dot \sigma \left[ - \frac{V_s}{\dot
\sigma^2} - b_\varphi \sin \theta \right] \, ,
\label{theta}
\ee
\be
V_\sigma = V_\varphi \cos \theta + e^{- b} \, V_\chi \sin \theta \, ,
\label{rotaz1}
\ee
and
\be
V_s = - V_\varphi \sin \theta + e^{- b} \, V_\chi \cos \theta \, .
\label{rotaz2}
\ee  

One of the main results of our work is that the coupling of 
adiabatic and isocurvature perturbations does not vanish on
super-Hubble scales {\em even} when
$\dot \theta = 0$. This is in contrast to the case of scalar fields with
ordinary coupling to gravity treated in Ref. \cite{Finelli:2001sr}, 
for which curvature and isocurvature perturbations
are decoupled when $\dot \theta = 0$. In particular, in the case of an
exponential potential for $\varphi$ treated in \cite{Brandenberger:2001bs}, 
our result implies
that curvature and isocurvature are coupled even if $\theta$ is constant.
Another way of reading this result is that adiabatic and isocurvature
perturbations are coupled by $V_s$: for scalar fields with canonical
kinetic terms,
$V_s$ is just given by $\dot \theta$, while when $ b_\varphi \ne 0$, $V_s$
contains extra terms, as we can see from Eq. (\ref{theta}).

We now introduce the Mukhanov variable \cite{mukhanov} $Q_\sigma$
related to $\delta \sigma$ by:
\be
Q_\sigma = \delta \sigma + \frac{\dot\sigma}{H} \Phi = \frac{\dot
\sigma}{H} \zeta \,.
\ee
The quantity $Q_\sigma$ is gauge-invariant and is used to quantize
cosmological perturbations \cite{mukhanov,Mukhanov:1990me}. It depends
both on the adiabatic component of the matter fluctuations, i.e. 
$\delta\sigma$, and on the metric fluctuations  $\Phi$. We can rewrite 
the equation of motion for the adiabatic perturbation mode as:
\begin{eqnarray} 
\ddot Q_\sigma &+& 3H \dot Q_\sigma  + \\
&& \left[ \frac{k^2}{a^2} + V_{\sigma \sigma}-\dot \theta^2 
- \frac{1}{M_{\rm pl}^2 a^3} 
\left(\frac{a^3  \dot\sigma^2}{H}\right)^{.} + b_\varphi u (t)
\right] Q_\sigma \nonumber
\\
= && 2 (\dot \theta \delta s)^. - 2 \left( \frac{V_\sigma}{\dot \sigma}
+ \frac{\dot H}{H} \right) \dot \theta \delta s \nonumber \\
&+& \left[ 2 b_\varphi h (t) 
+ b_{\varphi \varphi}\dot\sigma^2 \sin 2\theta\right]
\delta s \,,
\end{eqnarray}
where 
\be
V_{\sigma \sigma} =  V_{\phi \phi}(\cos \theta)^2 + V_{\phi \chi}
e^{-b} \sin 2\theta + V_{\chi \chi} (\sin \theta)^2 e^{-2b} \,,
\ee
\be
u (t) = - \dot \theta \dot \sigma \sin \theta - e^{-b} \, V_\chi \sin
\theta \cos \theta \, ,  
\ee
and
\be
h ( t ) = \dot \sigma 
(\sin\theta \delta s)^. 
- \sin \theta \left[\frac{\dot
H}{H} \dot\sigma + 2 V_\sigma \right] \delta s  - 3 H
\dot\sigma 
\sin \theta \delta s \, .
\ee
This equation reduces for $(b_\varphi = 0)$ to Eq.(55) in \cite{Gordon:2000hv}
in the case of two coupled scalar fields which both have canonical kinetic
terms.

We now differentiate Eq. (\ref{correct}) with respect to time
in order to get a second
order differential equation for $\zeta$:
\begin{eqnarray}
\ddot \zeta &+& (3 H - 2 \frac{\dot H}{H} + \frac{\ddot H}{\dot H}) 
\dot \zeta + \frac{k^2}{a^2} \zeta \label{zetaddot} \\ &=& \dot S + 
S (3 H - 2 \frac{\dot H}{H} + \frac{\ddot H}{\dot H} ) \nonumber \\
&=& \frac{H}{\dot \sigma} \left[ 
 (\dot \theta \delta s)^. - 2 \left( \frac{V_\sigma}{\dot \sigma}
+ \frac{\dot H}{H} \right) \dot \theta \delta s 
 \right. \nonumber \\
&+& \left. 
2 b_\varphi h (t)
+ b_{\varphi \varphi} \dot \sigma^2 \sin 2\theta\delta s
\right] \, . \nonumber
\end{eqnarray}

We note that for a power-law contraction $a(t) \sim (-t)^p$, 
the homogeneous part of the equation for $\zeta$ becomes:
\be
\ddot \zeta + 3 H \dot \zeta + \frac{k^2}{a^2} \zeta = 0 \,.
\ee
This is the equation for a standard 
massless minimally coupled scalar field, i.e.
the same equation which gravitational waves satisfy. As we have already
shown in \cite{Brandenberger:2001bs,Finelli:2001sr}, 
only a dust contraction ($p=2/3$) can generate a
scale invariant spectrum for the curvature perturbation. Therefore, even
allowing for the presence of a generalized free axion, we have a scale
invariant
spectrum for curvature perturbation only for the same type of contraction
already studied in the single field case, when the evolution of the scale 
factor is the same as dust.

In order to obtain the equation for the entropy field 
$\delta s$, we need to differentiate   
Eq. (\ref{entropyfield}) twice with respect to time and make use of Eqs. 
(\ref{deltaphi},\ref{deltachi}) as well as \cite{Notari:2002yc} 
\be
\epsilon_m = - \frac{k^2}{4 \pi G a^2} \Phi = \dot \sigma \dot \delta
\sigma - \dot \sigma^2 \Phi - \ddot \sigma \delta \sigma + 2 V_s \delta s
\,.
\ee
At the end we get:
\begin{eqnarray}
\ddot{\delta s} &+& 3 H \dot{\delta s} + \left[ \frac{k^2}{a^2} 
+ V_{ss} + 3 \dot \theta^2 - b_{\varphi \varphi} \dot \sigma^2 +
b_\varphi^2 g (t) + b_\varphi f (t) \right]
\delta s \nonumber \\
&=& - \frac{k^2}{a^2} \frac{\Phi}{2 \pi G} \frac{V_s}{\dot \sigma^2} 
\, .\label{entropy}
\end{eqnarray}
In the above, we have used the notation 
\begin{eqnarray}
g (t) &=& - \dot \sigma^2 (1 + 3 \sin^2\theta) \nonumber \\
f (t) &=& V_\varphi
(1 + \sin^2 \theta) - 4 V_s \sin \theta \, ,
\end{eqnarray}
where we have made use of the relations
(\ref{entropyfield},\ref{adiabaticfield},\ref{theta}-\ref{rotaz2})  
and the definition  
\be
V_{ss} =  V_{\phi \phi}(\sin \theta)^2 -  V_{\phi \chi} e^{-b}\sin 2\theta
+ V_{\chi \chi} (\cos \theta)^2 e^{-2b}.
\ee
We note that in the case of two coupled scalar fields which both have
canonical kinetic terms, 
$(b_\varphi = 0)$, Eq. (\ref{entropy}) reduces to Eq.(52) in
\cite{Gordon:2000hv}, but 
differs from Eq. (3.38) in \cite{Notari:2002yc}.

By using Eq. (\ref{correct}) to substitute the term $k^2 \Phi$, we can
rewrite Eq. (\ref{entropy}) as:
\begin{eqnarray}
\ddot{\delta s} +  3 H \dot{\delta s} &+& \left[ \frac{k^2}{a^2}
+ V_{ss} + 3 \dot \theta^2 +
b_\varphi^2 g (t) +
\right. \nonumber \\  
&+& \left. b_\varphi f (t)- b_{\varphi \varphi} \dot \sigma^2 - 4 \frac{V_s^2}{\dot
\sigma^2} \right] \delta s
= 2 \frac{V_s}{H} \dot \zeta \, .
\label{entropy2}
\end{eqnarray}

Eqs. (\ref{zetaddot}) and (\ref{entropy2}) 
represent the main result of this work. They 
describe the coupling between the adiabatic and the entropy fluctuation
modes. Eq. (\ref{zetaddot}) determines how the entropy fluctuation sources the
adiabatic mode, the Eq. (\ref{entropy2}) 
in turn gives the growth of the entropy mode
sourced by the adiabatic fluctuation component.

\section{An Exact Inflationary Solution}

As an application of the formalism developed in the previous section, an
application which is of interest in its own right, we
now consider a new inflationary solution based on the
action (\ref{action}) with 
\be
b (\varphi) = \frac{\alpha}{2 M_{\rm pl}} \varphi 
\ee
and with an exponential potential
\be
V = V_0 \, e^{- \beta \varphi/M_{\rm pl}} \,.
\label{potvarphi}
\ee
With this potential, the action (\ref{action}) is a model of {\em soft
inflation} \cite{Berkin:1990ju} with a constant potential 
for the inflaton $\chi$, i.e. $V(\chi) = V_0$.

The solution we are presenting is the expanding branch of the contracting
solution found in \cite{Finelli:2001sr}. We look for a solution 
for which the scale factor increases as a power of time, whereas the
inflaton depends logarithmically on time:
\begin{eqnarray} \label{ansatz2}
a(t) &\sim& t^{\, p} \,, \quad t>0 
\\
a(\eta) \, &=& \,(- \frac{1}{M_{\rm pl} (p-1) \eta})^\frac{p}{p-1} \,,
\quad \eta < 0
\\
\varphi (\eta)\, &=& \, A \log (- M_{\rm pl} (p-1) \eta) \label{ansatz1}
\,,
\end{eqnarray}
where we have written the time dependence of the scale factor both in
cosmic time $t$ and in conformal time $\eta$.

The solution for $\chi$ can be obtained by directly integrating 
Eq. (\ref{backchi}) with $V = V(\varphi)$. The resulting dependence
on $\eta$ is of power-law type:
\be
\chi ' = C \, \frac{e^{-\alpha \varphi/M_{\rm pl}}}{a^2} = 
\frac{C}{(- M_{\rm pl} (p-1) \eta)^{\alpha A/M_{\rm pl} - 2p/(p-1)}} \,,
\label{fform}
\ee
where $C$ is an integration constant and the prime denotes the derivative 
with respect to the conformal time. The ansatz (\ref{ansatz1}) and the
above functional form (\ref{fform}) for $\chi$ solve all of the
equations of motion provided certain relations between the constant
coefficients are satisfied. 
By imposing that all the terms have the same time dependence we get the
following constraints:
\begin{eqnarray} \label{constr2}
\beta \frac{A}{M_{\rm pl}} + \frac{2p}{p-1} &=& 2 \\   
\alpha \frac{A}{M_{\rm pl}} - \frac{4p}{p-1} &=& 2 \, ,
\end{eqnarray}
which leads to
\be
\alpha = \beta (1-3p) \, .
\label{alfabeta}
\ee
We also have 
\begin{eqnarray} \label{circle}
\left( \frac{A}{M_{\rm pl}} \right)^2 
+ \left( \frac{C}{M^2_{\rm pl}(p-1)} \right)^2 &=& \frac{2 p}{(p-1)^2} \,, \\
\frac{V_0}{M^4_{\rm pl}} &=& p (3p -1) \,.
\label{v0}
\end{eqnarray}
This solution has the property of having $\theta = \theta_0 = {\rm
const.}$ More precisely 
\begin{eqnarray}
\cos \theta = \frac{A (p-1)}{\sqrt{2p} M_{\rm pl}}  \quad &&
\sin \theta = \frac{C}{\sqrt{2p} M_{\rm pl}^2} \nonumber \\ 
\tan \theta &=& \frac{C}{A (p-1) M_{\rm pl}} \,.
\end{eqnarray}
The solution leads to inflation if $p > 1$. 
In Appendix B we show that the expanding solution is an
attractor when $p > 1/3$ and $\beta^2 > \beta_0^2$, where $\beta_0$ is
defined in Eq. (\ref{betarange}).

If both fields had standard kinetic terms, then curvature and
isocurvature perturbations would be decoupled since $\dot \theta = 0$
\cite{Gordon:2000hv}. Instead, for $\alpha \ne 0$, the isocurvature component 
feeds the curvature perturbation as described by Eq. (\ref{zetaddot}),
even in the long-wavelength limit ($k$ small). 
In addition, for $\alpha \ne 0$
curvature perturbations act as a source for isocurvature perturbations, as
we see from Eq. (\ref{entropy2}), but it is evident from Eq.
(\ref{entropy}) that this coupling is negligible for $k$ small.
Therefore, at least for small $k$, we can assume that isocurvature
perturbations evolve freely.

The homogeneous part of the solution of Eq. (\ref{zetaddot}) for the
curvature perturbation $\zeta$ is:
\be
\zeta_k = \frac{1}{2 a M_{\rm pl}} \sqrt{-\frac{\pi p \eta}{2}}
H_{|\nu_\zeta|}^{(2)} (-k\eta)
\,,
\label{zetasol}
\ee
where $H_{\nu}^{(2)}$ denotes the Hankel function of index $\nu$, and the
spectral index $\nu_{\zeta}$ is given by:
\be
\nu_\zeta = \frac{1}{2} \frac{3p-1}{p-1} = \frac{3}{2} + \frac{1}{p-1} \,.
\label{adindex}
\ee
For large cosmic time (very small negative conformal time), the Hankel
function scales as $(-k \eta)^{-\nu_{\zeta}}$. Hence, for $p > 1$,
the spectral index (\ref{adindex}) corresponds to a slightly blue
tilt away from a scale invariant spectrum, the tilt
decreasing as $p$ increases.

The left hand side of the equation for $\delta s$ (\ref{entropy})
in conformal time can be rewritten as:
\begin{eqnarray}
(a \delta s_k)'' &+& \left[ k^2 + a^2 \left( V_{ss} +
\frac{\alpha^2}{4 M_{\rm pl}^2} g(\eta) 
+ \frac{\alpha}{2 M_{\rm pl}} f(\eta) \right) \right. \nonumber \\ 
&& - \left. \frac{a''}{a} \right] (a \delta s_k) = 0 \,,
\end{eqnarray}
where 
\be
a^2 V_{ss} = a^2 (\sin\theta)^2 \, V_{\varphi \varphi} = 2 (3p-1)
\frac{(\tan\theta)^2}{(p-1)^2 \eta^2} \,,
\label{vsecond}
\ee 
\be
a^2 g(\eta) = - 2 p \frac{M_{\rm pl}^2}{\eta^2 (p-1)^2} (1 + 3
\sin^2 \theta) \,,
\ee
\be
a^2 f(\eta) = p \frac{\alpha M_{\rm pl}}{\eta^2
(p-1)^2} (1 + 5 \sin^2 \theta) \,,
\ee
\be
\frac{a''}{a} = p \frac{2p-1}{(p-1)^2 \eta^2} \,.
\ee
The solution for $\delta s$ is:
\be
a \delta s_k = \left( - \frac{\pi \eta}{4} \right)^{1/2} H_{|\nu_s|} (-k
\eta)
\ee
where $\nu_s$ can be obtained from:
\begin{eqnarray}
\nu_s^2 &=& \frac{1}{4} + p \frac{(2p - 1) + 6 (\tan \theta)^2
(1-3p)}{(p-1)^2} \nonumber \\ 
&=& \nu_\zeta^2 + 6 p (\tan\theta)^2 \frac{1-3p}{(p-1)^2}
\,.
\label{spectralindex} 
\end{eqnarray}

By comparing the index (\ref{adindex}) of the primordial adiabatic
fluctuations in this model (computed without the presence of an
entropy source) with the index (\ref{spectralindex}) of the entropy
fluctuation
mode, we see that in a model with potential (\ref{potvarphi})
the isocurvature perturbation spectrum (for $p > 1/3$) 
has a larger blue tilt than the adiabatic perturbation spectrum. 
Therefore, for such
power-law inflation models driven by a scalar matter field $\varphi$,
the presence of a free field $\chi$ 
does not change the basic picture of the dominance of 
adiabatic perturbations on large wavelengths, and can also not
change the index of the fluctuation power spectrum.

In the first two figures, the index $\nu_s$ is displayed as a function of $p$
and $\tan \theta$ in the case of the expanding Universe solution considered
in this section.

\begin{figure}[hpb]
\begin{tabular}{c}
\vspace{4cm}
\includegraphics{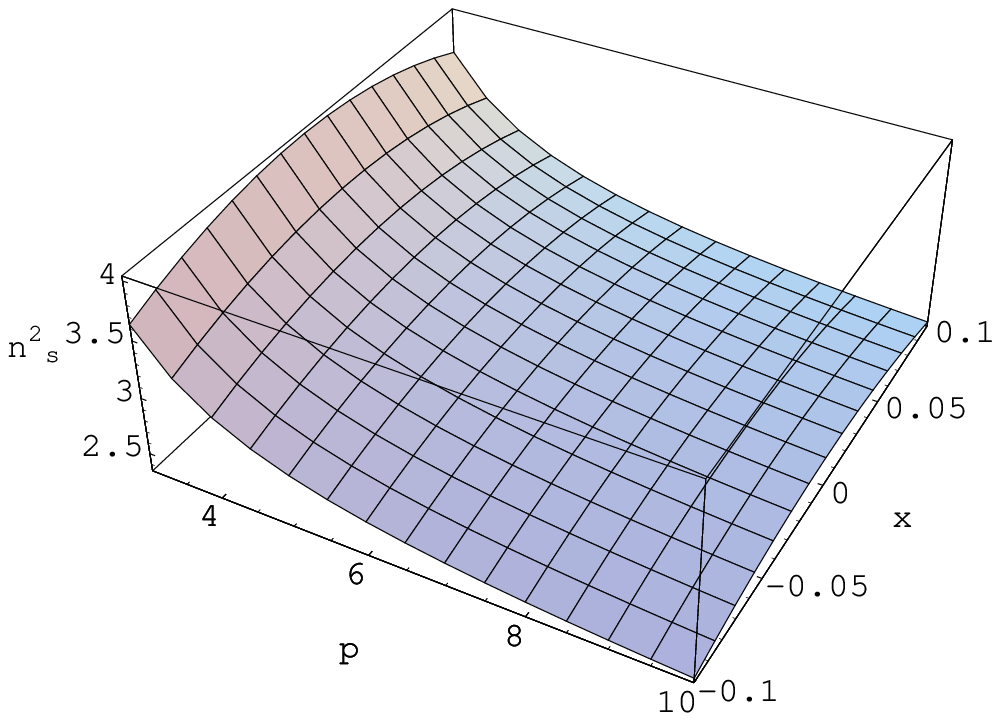}
\\
\vspace{4cm}
\includegraphics{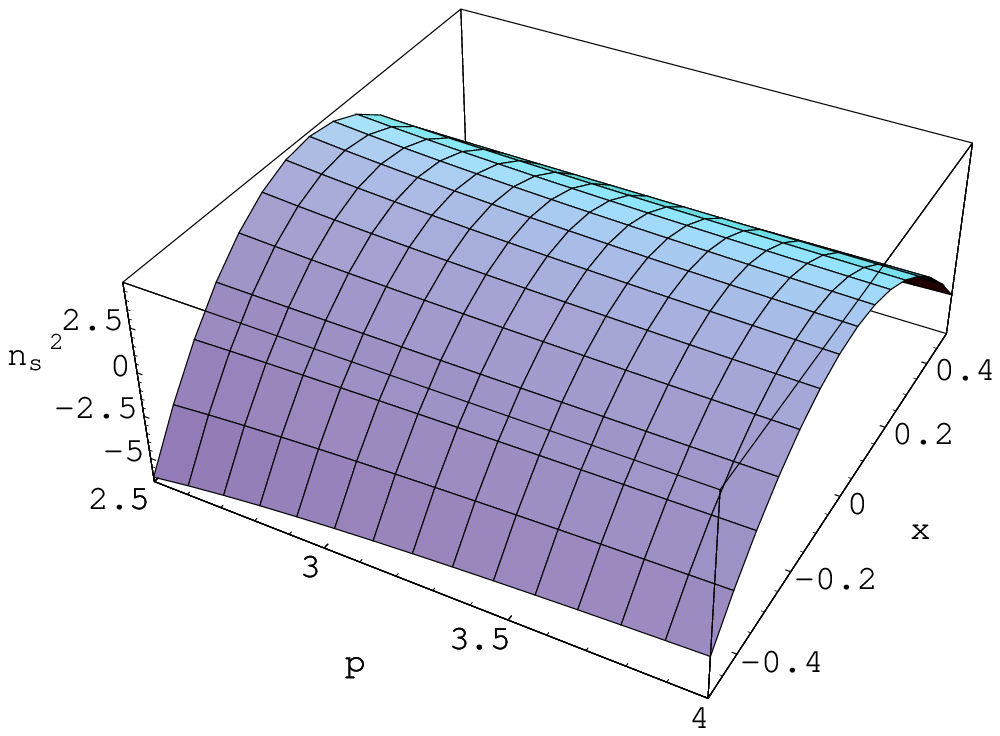}
\end{tabular}
\caption{The spectral index $\nu_s$ for isocurvature perturbations in
the expanding model of Section IV is presented as a function of 
$p$ and $\tan \theta$ (denoted by $x$ in the figure) for
two different ranges of the variables.} 
\label{exp}
\end{figure}

\section{The Contracting Solution}

We now discuss perturbations in the contracting branch of the
same two field model given by the potential (\ref{potvarphi}). 
In Appendix B we show that the contracting solution is not stable for
any value of the parameters $p$ and $\tan \theta$. However, the model
deserves attention since it allows an analytic study of
adiabatic and isocurvature perturbations. It also yields new ways
of obtaining a scale-invariant spectrum of fluctuations.

The contracting phase of this model could describe a modified PBB scenario in
which the dilaton has an exponential potential and the field $\chi$ denotes 
a generalized axion. Alternatively, $\chi$ could be
the five-brane position in an effective 4-D theory of the Ekpyrotic
scenario, with the four-form set to zero \cite{Copeland}.

Thus, here we consider a contracting background
(which would have the potential of solving the horizon problem):
\begin{eqnarray} \label{ansatz3}
a(t) &\sim& (-t)^{\, p} \,\,\,, t \, < \, 0 \,\,\,, 0 \, < \, p \, < \, 1  \\
a(\eta) \, &=& \,(- M_{\rm pl} (1-p) \eta)^\frac{p}{1-p} \,\,\,,
\eta < 0 \, ,
\end{eqnarray}
and the functional form of $\varphi(\eta)$ is the same as in the
expanding branch.

The formulae for the spectral indices 
$\nu_\zeta$ and $\nu_s$ of adiabatic and isocurvature
perturbations are the same as in the expanding case, but restricted to the
range $0 < p < 1$. For this range of $p$, isocurvature perturbations will
play an important role, as opposed to what happens in the case $p \gg 1$. 
Indeed, $|\nu_s| = 3/2$ occurs for the following set of values
of $\tan \theta$ 
\be
(\tan \theta)^2 = \frac{2-3p}{6 p (1-3p)} \, .
\label{isox}
\ee

Let us discuss the two limits in which the field $\chi$ is either 
subdominant or not. When $\chi$ is subdominant, $\tan \theta \sim 0$,
and $V_{ss}$ is
unimportant, as we can see from Eq. (\ref{vsecond}). In this
case, only the curvature term $a''/a$ can amplify isocurvature
perturbations, and therefore their spectrum is scale invariant only for
$p=2/3$ (since the equations are the same as in the case of 
adiabatic perturbations in the single field model 
\cite{Brandenberger:2001bs,Finelli:2001sr}). When the background $\chi$
energy is not negligible, a scale invariant spectrum of isocurvature
perturbations is possible for $0<p<1/3$ and $2/3<p<1$ (in order to have 
$(\tan \theta)^2 > 0$ in Eq. (\ref{isox})). 
When $\chi$ is
dominant, i.e. $\tan \theta \gg 1$, there are two values of $p$ which
yield a scale invariant spectrum of isocurvature fluctuations. The values
are obtained by setting the spectral index $\nu_s$ of 
(\ref{spectralindex}) to $\pm 3/2$. By combining (\ref{adindex}) and
(\ref{spectralindex}) it immediately follows that one of the two
solutions for $p$ is very close to $p = 0$ (but slightly positive, of the
order $(6 \tan^2 \theta)^{-1}$, the second solution has $p$ very
close to $p =  1/3$ (but slightly smaller). 
Note that these solutions yield new mechanisms of obtaining a
scale-invariant spectrum of fluctuations in a bouncing cosmology.
 
We now estimate the transfer of isocurvature perturbations to adiabatic
component. As has been recently understood, extra care should used,
since the usual decaying mode in the expanding case can become {\em
the} growing mode in the contracting case. Also, different quantities such
as $\Phi$ and $\zeta$, grow at different rates, as can
be seen from the long-wavelength solutions of $\zeta$ and $\Phi$ without
source terms (see e.g. \cite{Finelli:2001sr}):
\begin{eqnarray}
\zeta_k \, &=& \, D \, + \, S \int{{{d t} \frac{H^2}{a^3 \dot
\sigma^2}}}
\, , \\
\Phi_k \, &=& \, l \, D \, + \, m \, \frac{S}{k^2} \frac{H}{a} \,
\,,
\nonumber
\end{eqnarray}
where $D$ and $S$ are functions of $k$ and $l$ and $m$ are numerical
coefficients which depend on the equation of
state of the background.

Therefore, it is useful to check whether the growth of $\zeta$ is dominated
by the growth of the adiabatic mode which is driven by
the contraction of the universe, or by the contribution of isocurvature
perturbations. For this purpose, we estimate the two terms on the right
hand side of Eq. (\ref{correct}):
\be
\dot \zeta_k = k^2 p \beta \frac{S}{a^3} + \alpha \sin \theta
\frac{H}{M_{\rm pl}} \delta s_k \, ,
\ee
where we have considered only the relevant case of a growing mode for
$\Phi$ and of constant $\theta$. From this we can see that, 
for $p < 1/3$, the variation
in time of the adiabatic component is dominated in the long time limit by the
entropy contribution, even if we assume that $\delta s$ is constant.
Most likely also $\delta s$ grows in time. The typical transfer time is 
given by the Hubble parameter. From Eq. (\ref{entropy}), it follows that the
right-hand side is comparable at most to $\frac{S}{a^3}$.

Now, suppose we consider the case when the 
isocurvature perturbations have a scale invariant spectrum, 
given by one of the possible values of $\tan \theta$ of
Eq. (\ref{isox}). The solution for $\delta s$ at long-wavelengths ($-k
\eta \rightarrow 0$) is then given by:
\be
\delta s_k \sim
- i \frac{1-p}{\sqrt{2} k^{3/2} t} \, .
\ee
The curvature component induced by a scale invariant entropy component
will then grow in time in the same way, as can be seen from Eq.
(\ref{correct}):
\begin{eqnarray}
|\zeta_{k \, {\rm induced}}| 
&\sim& (1-p) \alpha \sin \theta \frac{H}{M_{\rm pl}} \frac{1}{\sqrt{2}
k^{3/2}} \nonumber \\
&\sim& \alpha \sin \theta p \frac{|\delta s_k|}{M_{\rm pl}}
\,.
\end{eqnarray}

This should be compared with the result for a scale invariant $\zeta$ in
the single field case simulating a dust collapse ($p = 2/3$) 
\cite{Finelli:2001sr}. In that case, for
long wavelengths:
\be
| \zeta_k | \sim \frac{H}{M_{\rm pl}} \frac{1}{ 18 \sqrt{6} k^{3/2}} \,.
\ee

\begin{figure}[hpb]
\begin{center}
\begin{tabular}{c}
\vspace{4cm}
\includegraphics{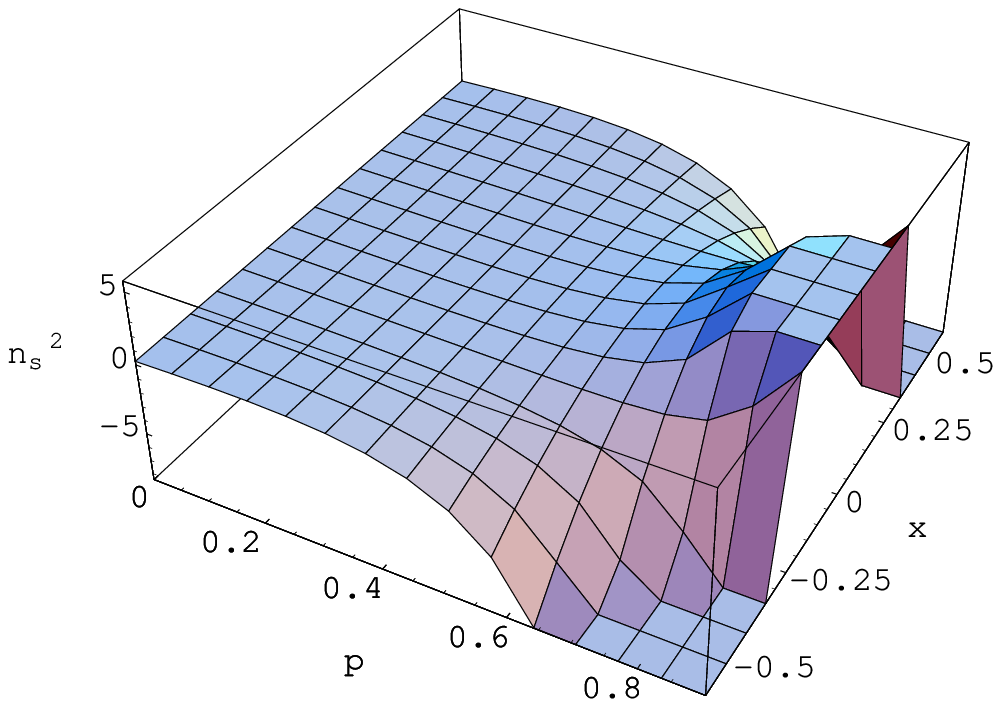} \\
\vspace{4cm}
\includegraphics{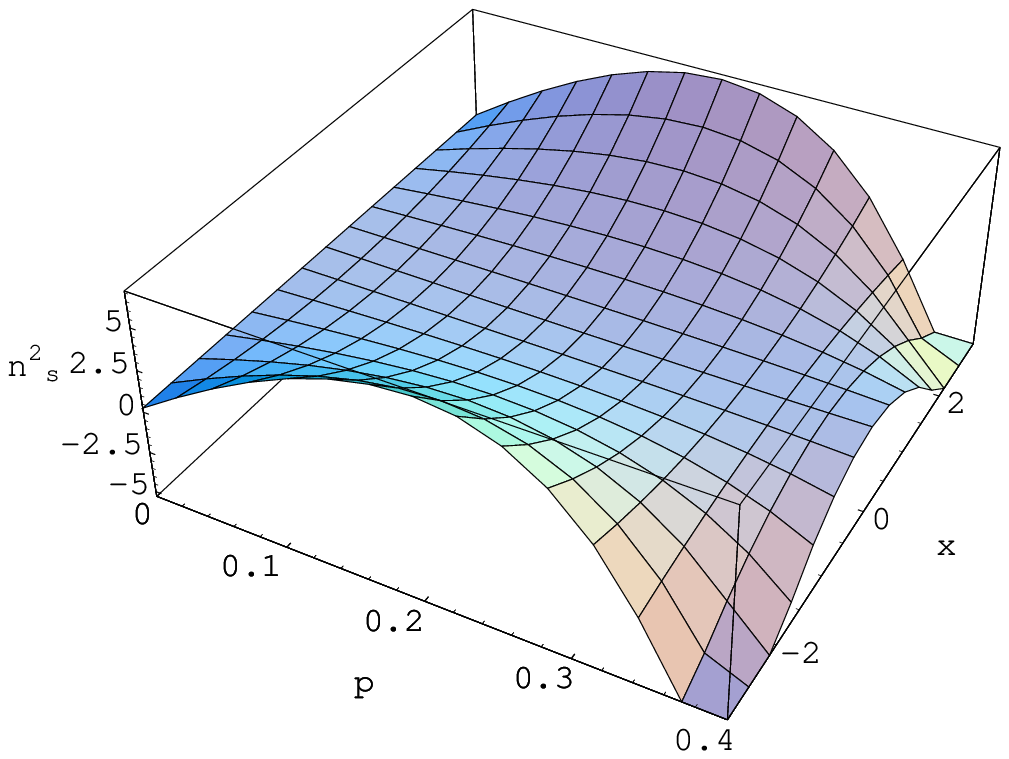}
\end{tabular}
\end{center}
\caption{The spectral index $\nu_s$ for isocurvature perturbations in
the contracting case is  
presented in function of $p$ and $\tan \theta$ (denoted by $x$) for
two different range of the variables.}
\label{contr}
\end{figure}

\section{Discussion and Conclusions}

We have studied the interaction between adiabatic and entropy 
perturbations for a
two field Lagrangian in which one field has a nontrivial kinetic term,
a prototypical effective scalar field model motivated by string theory.
We have thus extended the work by Gordon et al. \cite{Gordon:2000hv} to
a class of generalized Einstein theories likely to be relevant in
various approaches to string cosmology, be it an inflationary scenario
in which various moduli fields are dynamically important, or an
alternative scenario such as the PBB or Ekpyrotic paradigms.

We have discovered that the nontrivial kinetic term induces a
new coupling between adiabatic and entropy perturbations when a
non-zero potential term is present (see Eq. (\ref{correct})). This extra
coupling is not suppressed on super-Hubble scales, nor does it vanish in the
case of a scaling solution in which both fields participate in driving the
geometry (in contrast to what occurs for ordinary kinetic terms for a 
scalar field \cite{Malik:1998gy}). 
The reason why a transfer of power from the isocurvature mode to the
adiabatic mode can occur {\em even} for a scaling solution is because 
the two component are coupled ($b_\varphi \ne 0$), as also happens in
reheating \cite{reheating}, even at nonlinear level \cite{nonlinear}.

We have constructed specific models which yield a scale-invariant
spectrum of primordial isocurvature fluctuations (see Section V)
in a bouncing Universe scenario. Via the
coupling between entropy and adiabatic modes discussed in Section III,
a scale-invariant spectrum of adiabatic fluctuations in the post-bounce
phase is induced. However, these models are not stable, and
the specific values of the background fields required to obtain these
solutions do not appear natural. Thus, the question of using the
transfer mechanism discussed in this paper to construct improved
cosmological models is still open.

Models of the type studied here could be relevant in a framework where 
large-scale primordial adiabatic fluctuations are negligible and one 
therefore would like the late time adiabatic
perturbations to be seeded by the isocurvature component, 
see e.g. \cite{Linde:1996gt} for an early discussion in the context of
inflationary cosmology, or more recent discussions in the context of
the curvaton models \cite{Lyth:2001nq,Enqvist:2001zp,Moroi:2001ct}. 
In the conventional inflationary models, when both scalar fields have 
canonical kinetic terms, the curvature perturbation only
starts to grow at late times if the field carrying the isocurvature mode
has no nontrivial potential (as is the case for axion fluctuations
above the QCD scale), and the isocurvature component is converted 
fully to the adiabatic mode when the curvaton decays
\footnote{Note that
the background energy density of the curvaton must be negligible while its
perturbation are generated \cite{Linde:1996gt,Lyth:2001nq} or else there is no
potential to couple adiabatic and isocurvature perturbations 
\cite{Copeland:1997ug}.}.  
In contrast, in our models, the curvature fluctuation begins to grow 
early due to
the extra coupling between the entropy and the adiabatic mode 
\footnote{Note, however, that one still needs a mechanism to turn off the
growth of $\zeta$ on super-Hubble scales in order to produce
a spectrum which is adiabatic in the sense that the positions of the 
acoustic peaks in the induced CMB anisotropy
spectrum are at the values for a pure primordial adiabatic model.}. 
In this paper, we have studied examples in which the growth of $\zeta$ occurs
while the perturbations are generated. 
Moreover, when a scaling solution such as the one found in 
\cite{Finelli:2001sr} exists,
the conversion can be studied approximatively analytically.

The Lagrangian considered in Eq. (\ref{action}) is motivated by
low energy effective actions from string theory, often after
making use of a conformal trasformation. Such Lagrangians have
also been invoked to drive inflation; a partial
list of such models is given in 
\cite{Berkin:nm,Starobinsky:2001xq}. 
The splitting of adiabatic and isocurvature perturbation introduced in
this paper if therefore useful in this context. The expanding branch
of the exact solution studied in \cite{Finelli:2001sr} is relevant in 
the context of {\em soft inflation} models \cite{Berkin:1990ju} and we 
have demonstrated that it is an attractor (see also \cite{Felder:2002jk}).

In the context of string cosmology, the action in (\ref{action}) can be
obtained once one has compactified from ten to four dimensions and made
the transition to the Einstein frame. In this context,
the field $\chi$ could be the pseudo-scalar axion ($\alpha=2$ in units of
$M_{pl} = 1$) and $\varphi$ the shifted dilaton.
Alternatively, $\chi$ could be the modulus
of the rank-three internal antisymmetric tensor field and $\varphi$ the
modulus of the internal space ($\alpha = -2 / \sqrt{3}$ in units of
$M_{pl} = 1$) \cite{Lidsey:1999mc}.

In the context of the Ekpyrotic scenario \cite{Khoury:2001wf},
the case studied here is a simplification of the full 4D effective
Lagrangian \cite{Notari:2002yc,Copeland}. The field $\varphi$ 
could be the dilaton 
or the volume modulus and $\chi$ the five-brane position
\cite{Notari:2002yc,Copeland}. The introduction of a potential seems
fundamental since the free 4D effective action obtained from M-theory with
a moving brane \cite{Copeland} does not contain a scale invariant
perturbation, as we have shown in Appendix A. An alternative to the
introduction of a potential for the Ekpyrotic scenario is considering a
non vanishing four form and constructing a scale invariant perturbation as
in PBB, i.e. through an axion \cite{Copeland:1997ug}.

In Appendix B we have demonstrated that the novel scaling solution found
in \cite{Finelli:2001sr} is a global attractor for $p>1/3$ and 
$\beta>\beta_0$. The
solution leading to inflation is therefore stable. In contrast, the
contracting scaling solution is never stable 
\cite{Felder:2002jk,Heard:2002dr}. 
This means that the scaling single field
contracting solution used in the Ekpyrotic scenario or a modified
PBB scenario (and demonstrated to be stable in \cite{Heard:2002dr}) is not
robust to the introduction of a second field with a nontrivial kinetic
term (it does not matter if this second field corresponds to 
an axion or something else).

We wish to end with a brief discussion of the prospect of using the
coupling between entropy and adiabatic fluctuation modes discussed
in this paper in order to obtain a scale-invariant spectrum of
adiabatic fluctuations in PBB and Ekpyrotic type models in which
the primordial adiabatic spectrum is not scale-invariant. The idea
is to use the coupling discussed in this paper in order to transfer
an initial isocurvature mode to the adiabatic component during the
phase of cosmological contraction (when viewed from the Einstein
frame). The induced spectrum for $\zeta$ will be scale-invariant for
the appropriate value of the spectral index $\nu_s$ 
(see (\ref{spectralindex})). After matching across the bounce
to a conventional expanding Friedmann cosmology (with no source
of entropy fluctuations) using
the method of \cite{Brandenberger:2001bs}, the spectrum of the
constant mode of $\zeta$ in the expanding phase
will be scale-invariant, unless the presence of the field $\chi$
modifies the matching conditions in an unexpected way (the study
of matching conditions in such a two-field model is currently under way).
Thus, if the transfer of an initial isocurvature fluctuation 
into a curvature fluctuation takes place in the contracting phase, one
does not need to introduce ad hoc new physics in the expanding phase
(as has to be done in the {\it curvaton} models) to
turn on the entropy source.

\vspace{1cm}

{\bf Acknowledgments}

F. F. is grateful to David Wands for useful correspondence and discussions 
on the phase space analysis and to Andre Lukas for useful discussions.
This work was supported in part (at Brown) by the U.S. Department 
of Energy under Contract DE-FG02-91ER40688, TASK A,

\section{Appendix A: Perspective on Ekpyrotic Models without Moduli Potentials}

In this Appendix we demonstrate that without a potential for some of
the moduli fields, it is not possible to obtain a scale-invariant spectrum
of fluctuations in effective field theories containing the moduli fields
which are expected to arise in the original Ekpyrotic model (in which
the separation between the boundary branes is fixed and a bulk brane is
propagating) \cite{Copeland}. In this model, isocurvature mode arise
naturally
since three scalar fields are present in the low-energy 
four-dimensional effective action.

Following 
Copeland, Gray and Lukas \cite{Copeland}, the kinetic part of the action is: 
\begin{eqnarray}
S_{eff} &=& \int d^4 x \sqrt{-g} \left[ - \frac{1}{4}
\partial_\mu \phi \partial^\mu \phi - \frac{3}{4}
\partial_\mu \beta \partial^\mu \beta \right. \nonumber \\
&& \left. - \frac{c}{2} e^\frac{\beta - \phi}{M_{\rm pl}}
\partial_\mu z \partial^\mu z \right] \,
\label{effective}
\end{eqnarray}
where $\phi$ is the dilaton, $\beta$ is the volume modulus and $z$
the brane modulus \cite{Copeland}.

It is interesting to consider the resulting cosmological perturbations
arising from the above theory when no potential is present. Since there is
no potential, adiabatic and isocurvature perturbations (in this case there
are two isocurvature modes) are decoupled. Instead of proceding by brute
force by considering perturbations for the theory (\ref{effective}), we
follow the method of \cite{Copeland}.

As observed in \cite{Copeland}, the action (\ref{effective}) is manifestly
invariant under a global SL(2,$R$) transformation, as can be seen
by defining:
\be
X = \sqrt{\frac{3}{8}} \beta - \sqrt{\frac{3}{8}} \phi \,, 
\ee
\be
Y = \frac{3}{\sqrt{8}} \beta + \frac{1}{\sqrt{8}} \phi \,. 
\ee
In terms of these fields the action (\ref{effective}) takes the form:
\begin{eqnarray}
S_{eff} &=& \int d^4 x \sqrt{-g} \left[ - \frac{1}{2}
\partial_\mu X \partial^\mu X - \frac{1}{2}
\partial_\mu Y \partial^\mu Y \right. \nonumber \\
&& \left. - \frac{1}{2} e^{2 \sqrt{\frac{2}{3}} \frac{X}{M_{\rm pl}}}
\partial_\mu z \partial^\mu z \right] \,.
\label{effective2}
\end{eqnarray}
where we have absorbed $c$ in the redefinition of $z$. In this way, $Y$
decouples and $X$ and $z$ parametrize the SL(2,$R$)/U(1) coset. 
This symmetry group has been already extensively used in the context 
of PBB cosmology
\cite{Copeland:1997ug}. In fact, the action is the same as the one
considered in \cite{Lidsey:1999mc}. Here, the pseudo-scalar axion field
$\sigma$ of
the rank-three antisymmetric strength field tensor is replaced by the bulk
brane position $z$.

In order to calculate the isocurvature perturbations induced by the bulk
brane, we follow the analysis used in PBB cosmology
\cite{Copeland:1997ug,Lidsey:1999mc}.
We know that both background and
perturbation solutions with $z' \ne 0$ are generated by acting with a
symmetry group transformation on the background and
perturbation solutions with $z' = 0$. It is thus sufficient to solve the case
with $z' = 0$ in order to derive the spectral index of the perturbations. 

For $z'=0$ the bulk brane perturbations satisfy the following equation:
\be
\ddot{\delta z}  + \left( 3 H + 
2 \sqrt{\frac{2}{3}} \frac{\dot X}{M_{\rm pl}} \right) \dot{\delta z} +
\frac{k^2}{a^2}
\delta z = 0
\ee
The solution for the scale factor is $a (t) \sim t^{1/3}$, and the
fields depend logarithmically on time (we are omitting integration constants):
\begin{eqnarray}
X &=& \, M_{\rm pl} r \log (- M_{\rm pl} \eta) \nonumber \\
Y &=& \, M_{\rm pl} s \log (- M_{\rm pl} \eta) 
\end{eqnarray}
where the coefficients $r$ and $s$ are subject to the constraint 
\be
r^2 + s^2 = 3/2 
\label{constraints}
\ee
The solutions for the bulk brane fluctuations are:
\be            
\delta z = e^{-\frac{2}{3} \frac{X}{M_{\rm pl}}} H_{|\nu_z|} (-k\eta)
\ee
where 
\be
\nu_z = \sqrt{\frac{2}{3}} r \,.
\ee
Because of the constraint (\ref{constraints}), the spectral index
$\nu_z$ can be at most $1$ (when $y$ is constant), and therefore,
in the context of this 4D general effective action for low-energy
heterotic M theory, it is not possible to obtain a
scale-invariant spectrum of fluctuations without adding a superpotential
or considering a nonzero four form. 

\section{Appendix B: Stability Analysis}

Here we study the stability of the scaling solution found in
\cite{Finelli:2001sr} and discussed in Section 4 and 5. 
In the context of exponential potentials 
for scalar fields,
other scaling solutions exist: for the system
perfect fluid-scalar field a scaling solution exists \cite{Wetterich:fm} and
has been shown to be a global attractor \cite{Wands:zm}. A scaling
solution exists even if the scalar field is explicitly coupled with a
perfect fluid \cite{coupled}. 
A scaling solution with multiple
fields and generic exponential potentials is relevant in the context of
assisted inflation \cite{assisted} and contraction \cite{assicontr}. The
accelerating expanding branch is stable \cite{Malik:1998gy} as in the single
field case \cite{Halliwell:1986ja}. Recently it has been shown that also the
single field contracting solution with a negative exponential is stable
\cite{Heard:2002dr}. 

We now introduce 
\be
x = \frac{\dot \phi}{\sqrt{6} H M_{\rm pl}} \quad 
y = \frac{\sqrt{|V|}}{\sqrt{3} H M_{\rm pl}} \quad w = \frac{e^b
\dot \chi}{\sqrt{6} H M_{\rm pl}}
\label{xyz}
\ee
and the system of equation of motion of the scalar fields can be written
as:
\begin{eqnarray} \label{system}
x' &=& -3 x \pm \beta \sqrt{\frac{3}{2}} y^2 + \alpha \sqrt{\frac{3}{2}}
w^2 + 3 x (x^2 + w^2) \nonumber \\
y' &=& y \left( - \beta \sqrt{\frac{3}{2}} x + 3 (x^2 + w^2) \right) \\
w' &=& w \left( - 3 - \sqrt{\frac{3}{2}} \alpha x + 3 (x^2 +w^2) 
\right) \nonumber
\end{eqnarray}
where a prime denotes a derivative with respect to the logarithm of the
scale factor. The Hubble law constrains the three variables to stay on a
circle/hyperboloid 
\be
x^2 \pm y^2 + w^2 = 1 \,,
\label{constraint}
\ee
depending on the sign of the potential. We have considered the absolute
value of $V$ in the definition of $y$ in Eq. (\ref{xyz}) and the
plus/minus signs in Eqs. (\ref{system},\ref{constraint}) takes into
account the possibility of having
positive or negative potentials, as in \cite{Heard:2002dr}.
We note that $y=0$ and $w=0$ are invariant planes of the dynamics. 

Following Ref. \cite{Heard:2002dr}, we restrict our study of the existence
and stability of critical points to the region $y \ge 0$, i.e. expanding
cosmologies ($H > 0$). However, the trajectories are symmetric with
respect to time reversal $H \rightarrow - H$, corresponding to $x
\rightarrow x$, $y \rightarrow - y$ and $w \rightarrow w$. The system is
also symmetric under $\beta \rightarrow - \beta$, $\alpha \rightarrow -
\alpha$ and $x \rightarrow - x$, and thus we restrict our attention
to the case with $\beta > 0$.

The system admits several fixed
points: \hfil\break
a) for $w=0$ the system has the fixed points of the single field
problem \cite{Heard:2002dr}.\hfil\break
b) For $y=0$ the system reduces to a
generalized axion-massless scalar field problem and there are no fixed
points, except $x=0$, $y=0$ (indeed, the scaling solution found in
\cite{Finelli:2001sr} does not exist for zero potential). 
We also note that the free axion-dilaton solution \cite{clw} is not a
scaling solution. \hfil\break
c) In the general case, the fixed point
\be
x = \frac{\sqrt{6}}{\beta - \alpha} \quad 
y = \pm \sqrt{\frac{\alpha}{\alpha - \beta}} \quad 
w = \pm \sqrt{\frac{1}{\beta - \alpha} \left( \beta -
\frac{6}{\beta - \alpha} \right)} 
\label{fixedpoint}
\ee
exists and corresponds to the solution presented in Sections 4 and 5 and in
Ref. \cite{Finelli:2001sr}. In the following, we use Eq. (\ref{alfabeta}) 
to substitute $\alpha$ with $\beta$ and $p$.

\begin{figure}
\epsfxsize=2.9 in \epsfbox{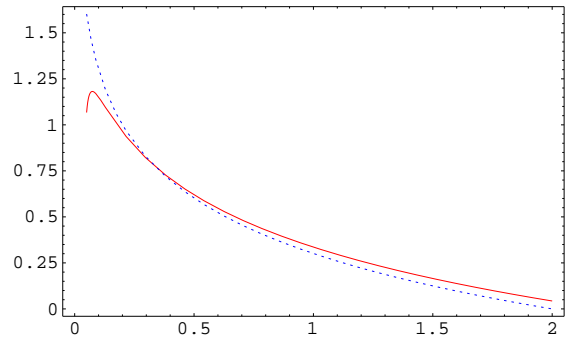}
\caption{The functions $\beta_0^2 (p)$ (continuous line) and $2/p$ (dashed
line). The two functions intersect in $p=1/3$, and $\beta_0^2 > 2/p$ for
$p>1/3$.}
\label{topo}
\end{figure}

By linearizing the system (\ref{system}) around the fixed point,
(\ref{fixedpoint}) we find two eigenvalues \footnote{A third eigenvalue,
$\lambda_3 = \frac{2}{p}$, with its relative eigenvector, does not
satisfy the linearization of the constraint (\ref{constraint}) and
therefore must be rejected. We thank David Wands for making this point
clear.}: 
\be
\lambda_{1,2} = \frac{1 - 3p \pm \sqrt{1 - 30 p + 81 p^2 + 12 \beta^2 p^2
- 36 \beta^2 p^3}}{2p} \,.
\ee 
The two eigenvalues are complex conjugates when $p>1/3$ and $\beta^2 >
\beta_0^2$ or when $p<1/3$ and $\beta^2 < \beta_0^2$, where
$\beta_0$ is given by:
\be
\beta_0^2 \equiv \frac{81 p^2 - 30 p + 1}{12 p^2 (3 p - 1)} =
\frac{27 p -1}{12 p^2} \,.
\label{betarange}
\ee
We also have the constraint $\beta^2 > 2/p$. 
We show the curves for $\beta_0^2 (p)$ and
$2/p$ in Fig. (\ref{topo}). For $p>1/3$ and $\beta^2 > \beta^2_0$ the
real part of the two eigenvalues is negative: in this case the solution found
in \cite{Finelli:2001sr} is an attractor in the expanding case, including the
inflationary case ($p>1$). For $p>1/3$ and $\beta^2 < \beta^2_0$ the two
eigenvalues are real, but have opposite signs, and therefore the fixed
point is a saddle point. 
For $p<1/3$, $\beta_0^2 < 2/p$ and therefore the two eigenvales are real,
with opposite signs. Also in this last case the fixed point is a saddle
point. 

Because of the symmetry under time reversal, early time solutions in an
expanding universe are the same as late time ones in a contracting
universe \cite{Heard:2002dr}. Therefore, the solution in the contracting
case is never stable: this means that the introduction of a generalized
free axion destroys the stability of the single field contracting
scaling solution
with negative potential ($p<1/3$) studied in \cite{Heard:2002dr}.   

\vspace{2cm}


\end{document}